\documentclass[10pt,twocolumn,twoside]{article}
\usepackage{graphicx} 

\usepackage{siunitx}
\usepackage{physics}
\usepackage{authblk}
\usepackage[normalem]{ulem}

\newcommand{\corr}[1]{\textcolor{black}{#1}}

\title{Asymmetric Bending Boundary Layer: the $\lambda$-test}
\author[a]{Nathan Vani}
\author[a]{Alejandro Ibarra}
\author[a]{José Bico}
\author[a]{Étienne Reyssat}
\author[a, b]{Benoît Roman}

\affil[a]{PMMH, CNRS UMR-7636, ESPCI Paris, PSL Université, Sorbonne Université, Université Paris Cité, Paris 75005, France}
\affil[b]{Corresponding author: benoit.roman@espci.fr}

\begin{document}

\maketitle

\begin{abstract}
We investigate the mechanics of two asymmetric ribbons bound at one end and pulled apart at the other ends. We characterize the elastic junction near the bonding and conceptualize it as a bending boundary layer. While the size of this junction decreases with the pulling force, we observe the surprising existence of the binding angle as a macroscopic signature of the bending stiffnesses. Our results thus challenge the standard assumption of neglecting bending stiffness of thin shells at large tensile loading. In addition, we show how the rotational response of the structure exhibits a non-linear and universal behavior regardless of the ratio of asymmetry. Leveraging the independence of the binding angle to the pulling force, we finally introduce the $\lambda$-test — a visual measurement technique to characterize membranes through simple mechanical coupling.
\end{abstract}

\paragraph{Significance statement.} It is commonly assumed that the bending stiffness of slender structures can be neglected under large tensile stresses. Our study reveals a critical exception at the junction of two bound ribbons pulled apart, where a scale-invariant binding angle emerges as a macroscopic geometric marker of the stiffnesses. The elastic junction is furthermore geometrically and mechanically characterized as a boundary layer. This boundary layer, ubiquitous in everyday objects featuring coupled slender structures, can be used to devise a remarkably straightforward visual method for measuring the stiffness of membranes.

\paragraph{Introduction.} The coupling of deformable solids is ubiquitous in engineering as well as in biology, with junctions often being structural weak points \cite{timoshenko1965theory, lecuit2007cell, leger2008adhesion}. While research on adhesion and assembly has so far largely covered the mechanical behavior of adhesive layers \cite{bigoni1997effect, maugis2013contact, creton2016fracture}, systems of coupled slender bodies have received comparatively little attention. Existing studies indeed primarily focus on the rich static \cite{vandeparre2011wrinkling, bigoni2011experimental, lechenault2014mechanical, cazzolli2020elastica, radisson2023elastic} and dynamic \cite{audoly2005fragmentation, callan2012self, kodio2020dynamic} behavior of a single object. We are interested here in a common yet understudied geometric feature: the seams that appear when the extremities of two elastic plates are glued flat together. Concerned systems include the classical T-peel test used for measuring adhesion energy \cite{kaelble1959, moore2003determination, bartlett2023peel}, tearing \cite{anderson2005fracture, takei2013}, heat-sealed Mylar balloons such as mundane potato chips bags \cite{paulsen1994shape} and tubular inflatables \cite{siefert2019programming}, flexible architected materials \cite{bertoldi2017flexible, tricard2020ribbed}, \textit{kirigami}~\cite{yang2018multistable} and \textit{kuttsukigami}~\cite{twohig2024kuttsukigami} structures, as well as the familiar shape of a book spread open. In all of these objects, the loading of the coupled plates results in significant tension near the seam.

\begin{figure*}
\centering
\includegraphics[width=1\textwidth]{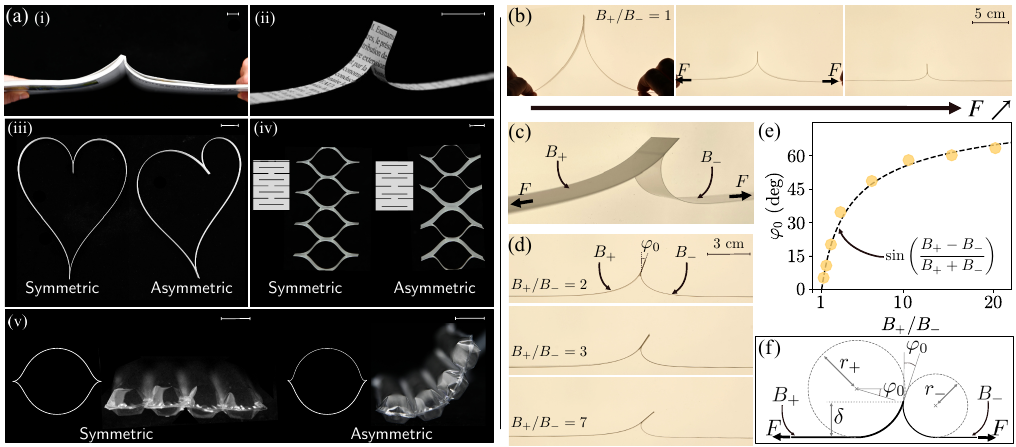}
\caption{Elastic junctions and their binding angle. (a) Illustration of the bending boundary layer in various systems: (i) Holding a magazine, (ii) Tearing a strip of newspaper, (iii) \textit{Kuttsukigamis} in the form of a heart \cite{twohig2024kuttsukigami} obtained by joining identical ribbons (left) and asymmetric ribbons (right), (iv) planar \textit{kirigamis} with symmetrical (left) and asymmetrical (right) unit cells along a top view of their cutting pattern, (v) Mylar tubes in series made from soldering two symmetrical (left) or asymmetrical (right) thin sheets along parallel seam lines. Though both cross-sections are mostly circular, the bending boundary layer appears on each side and control the overall curvature. Scale bars are $1$ \si{\cm}. (b) Sequential pictures of bound ribbons pulled apart in the case of equal stiffness ($B_+/B_- = 1$). (c) 
Pair of coupled ribbons of width $1.5$ \si{\cm}. (d) Pictures of the junction region observed for increasing ratios of bending stiffness ($B_+/B_- = 2,$ $3,$ $ 7$). (e) Measured junction angle $\varphi_0$ versus the ratio of bending stiffness. The dashed line corresponds to Eq.~\ref{eq:main_sin}. (f) Geometric approximation of the system.}
\label{fig:introduction}
\end{figure*}

Fig.~\ref{fig:introduction} (a) exhibits some of those systems involving coupled slender elements. Our research is motivated by the presence of a shared feature across these objects: an elastic junction where two slender beams are bound tangentially and form a cusp. This geometrical coupling occurs for instance when two elements are stuck together (i, iii and iv) or when the elements themselves are created as the result of a partial cut in a sheet (ii and iv). To characterize this phenomenon, we focus on the behavior of a minimal system made of two bound ribbons.

Fig.~\ref{fig:introduction} (b) and (c) illustrate this model structure. Two elastic ribbons are taped together at one end and pulled apart in opposite directions at their free ends (see Materials and Methods). As the pulling force is increased, most of the ribbons become straight and the extension of the bent regions near the junction tends to diminish. Unexpectedly, our research shows that the bending stiffnesses yield a macroscopic geometric feature even under substantial loading. We describe the geometry of this elastic junction as a function of the loading with an emphasis on the non-symmetric case. A boundary layer is identified and characterized. We then turn to the rotational response of the structure which also exhibits an interesting non-linear universal behavior. Finally, we introduce the ``$\lambda$-test", a  novel method using our findings to measure bending stiffness, named after the characteristic shape of the boundary layer. We believe this versatile test to be of direct use to experimental work in numerous fields of mechanics and soft matter as stiffness is particularly difficult to measure for thin sheets.

\paragraph{Binding angle.} When both ribbons are identical, the system exhibits a \corr{reflective symmetry about the cusp} and the joint part points normal to the pulling direction (Fig.~\ref{fig:introduction} (b)). In contrast, coupling ribbons of different stiffness breaks this symmetry and introduces a binding angle $\varphi_0$ between the connected ends and the normal to the pulling direction. Fig.~\ref{fig:introduction} (d) shows three examples with increasing stiffness asymmetry. A striking observation in Movies S1-2 is the independence of the angle $\varphi_0$ from the pulling force. Regardless of the asymmetry, the ribbons deform in-plane only with no significant stretching of the material. We thus describe each ribbon by a planar bending rod, referred to since Euler as  \textit{elastica}~\cite{euler1744methodus}. Noting the respective bending stiffnesses of the ribbons $B_+$ and $B_-$ with $B_+/B_- \geq 1$, the coupled system obeys the following equations:
\corr{
\begin{equation}
        \begin{cases}
        \begin{aligned}
            B_+\dfrac{\dd^2 \theta_+}{\dd s_+^2} - F\sin{\theta_+} = 0, \\
            B_-\dfrac{\dd^2 \theta_-}{\dd s_-^2} - F\sin{\theta_-} = 0,
        \end{aligned}
        \end{cases}
        \label{eq:def_system}
\end{equation}
}

\noindent\corr{where $F$ is the norm of the pulling force. For the sake of symmetry in these equations, arc lengths $s_+$ and $s_-$ and local angles $\theta_+$ and $\theta_-$ of the tangent along the rods are oriented symmetrically, increasing towards the junction point (see also Fig S1 in SI). We thereafter denote $\dot{\theta}_+$ and $\dot{\theta}_-$ as the derivatives of the angles with respect to $s_+$ and $s_-$, respectively.} The two conditions at the connection point are the balance of torques $B_+\dot{\theta}_+ = B_-\dot{\theta}_-$ and geometric compatibility $\theta_+ = \pi - \theta_-$, with $\varphi_0 = \pi/2 - \theta_+$. The asymptotic pulling at the free ends imposes both the angles and their derivatives to be null at an infinite distance from the junction. Integration of Eqs.~\ref{eq:def_system} at the junction gives \corr{$2FB_+ \left(\sin\varphi_0 - 1 \right) + 2FB_- \left(\sin\varphi_0 + 1 \right) = 0$} and the expression for $\varphi_0$ (see SI Appendix for full derivations):
\begin{equation}
    \sin \varphi_0 = \frac{B_+ - B_-}{B_+ + B_-}.
  \label{eq:main_sin}  
\end{equation}

We do find that the junction angle $\varphi_0$ depends only on the stiffness asymmetry $B_+/B_-$, as well as a good quantitative agreement with experiments in Fig.~\ref{fig:introduction} (e). A physical interpretation is obtained by considering a simplified system in which the ribbons are represented by two aligned straight lines connected by arcs of circles tangent at the junction with radii $r_+$ and $r_-$, as in Fig.~\ref{fig:introduction} (f). Geometrically, the height $\delta$ of the junction reads $\delta = r_+\left( 1 - \sin\varphi_0\right) = r_-\left( 1 + \sin\varphi_0\right)$, so that the angle $\sin \varphi_0 = (r_+ - r_-)/(r_+ + r_-)$ is a geometric expression of the curvature asymmetry, which in turn derives solely from torque continuity at the junction $B_+/r_+ = B_-/r_-$. This leads to Eq.~\ref{eq:main_sin}, with $\varphi_0$ independent of the pulling force. Flexible structures generally tend to align along large tensile forces $F$ except in a region of characteristic length scale $\ell$ obtained by balancing the bending torque $B/\ell$ with the moment $F\ell$, leading to $\ell \sim \sqrt{B/F}$ \cite{kaelble1959}. As $\ell$ vanishes for large loads, we usually ignore the bending stiffness. In contrast, we find here a constant deflection angle $\varphi_0$ that remains a strong geometric marker of the bending stiffness ratio $B_+/B_-$ which cannot be discarded.

The example of the Mylar balloon, illustrated in Fig.~\ref{fig:introduction} (a) (v), perhaps most compellingly demonstrates the potential significance of the binding angle. Regardless of the stiffness asymmetry, the cross-section of the inextensible tubes substantially pressurized can be approximated by a circle. This result stems from the common assumption of neglecting bending stiffness, leading to volume maximization \cite{paulsen1994shape, skouras2013computational, siefert2019programming, panetta2021computational}. Yet, it is the connecting seam lines that determine the relative orientation of consecutive parallel tubes. The binding angles on each side of the tubes thus govern the rotation. Even a slight stiffness asymmetry can lead to a strong curvature of the global structure perpendicular to the tubes, as seen in the example on the right with a stiffness asymmetry of $B_+/B_- \sim 2$. Such boundary layers could potentially be used as hinge-like mechanisms to control the curvature of large-scale objects through the strategic arrangement of connections between sheets.

\begin{figure}[tbhp]
\centering
\includegraphics[width=\linewidth]{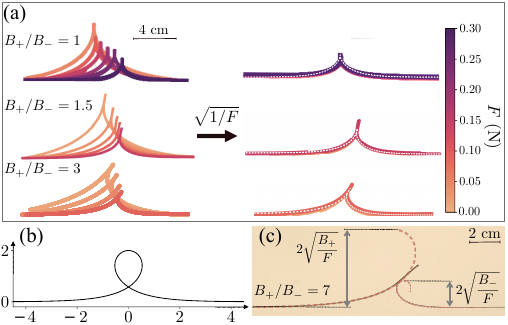}
\caption{Scaling and shape of the junction. (a) Extracted profiles of coupled ribbons as the pulling force $F$ is increased for different ratios of bending stiffness (left), and the same shapes rescaled by $\sqrt{1/F}$ showing good collapse (right). White dashed lines are the predicted shapes as obtained through graphical construction using the \textit{elastica} solution Eq. \ref{eq:full_sol_loop}. (b) Aforementioned solution plotted in normalized spatial units. (c) Coupled ribbons ($B_+/B_- = 7$) pulled apart. The dashed curves correspond to Eq. \ref{eq:full_sol_loop}.} 
\label{fig:geometry}
\end{figure}

\paragraph{Shape of the boundary layer.} Fig.~\ref{fig:geometry} (a) shows the profiles of three samples pulled with increasing force. The characteristic radii of curvatures of the ribbons being respectively $\sqrt{B_+/F}$ and $\sqrt{B_-/F}$, we obtain an overall collapse of the profiles at a given ratio $B_+/B_-$ when rescaled by $\sqrt{1/F}$. The reported self-similar behavior holds for ribbons that are long enough relative to their characteristic curvature. Under this condition, a family of curves parameterized by $B_+/B_-$ universally describes the shape of the junction of coupled slender structures under tension.

The curve described by the junction corresponds to the recombination of two parts of a classical solution of \textit{elastica}, more precisely the one of a rod pulled asymptotically and presenting a single loop (solution class number 7 in Euler seminal work) \cite{euler1744methodus, truesdell1960rational}. \corr{Fig.~\ref{fig:geometry} (b) shows this solution, expressed in curvilinear coordinates (with the arc length orientated from left to right and with $s=0$ corresponding to the apex of the loop) as \cite{landau2013fluid}:}
\corr{\begin{equation}
    \cos \frac{\theta\left(s\right)}{2} = -\tanh \left( s\, \sqrt{\frac{F}{B}}\right).
    \label{eq:full_sol_loop}
\end{equation}}

\noindent The shape of each ribbon follows Eq. \ref{eq:full_sol_loop}, with their respective scales $\sqrt{B_+/F}$ and $\sqrt{B_-/F}$, up to the junction. The junction shape is thus obtained by aligning vertically the baselines of two properly scaled solutions, and finding the first point of contact when approaching horizontally the curves (shown in SI Appendix F). Fig.~\ref{fig:geometry} shows the predicted shape overlaid (a) on the collapsed profiles and (c) on an experimental picture with $B_+/B_- = 7$, both with good agreement.

Drawing a brief parallel to wetting, one notes that the family of shapes described above is the solution to another related problem. The bending rod equation as in Eqs.~\ref{eq:def_system} also dictates the shape of a two-dimensional liquid meniscus \cite{landau2013fluid, roman2020pendulums}. Consider then a stiff, freely rotating plate that separates two completely wetting meniscii associated with different surface tensions $\gamma_+$ and $\gamma_-$ in a vast container reminiscent of a Langmuir-Blodgett trough (as illustrated in SI Fig. S8). The global torque balance of the plate would impose that the meniscii on each side of the plate should match in height at their triple point. The geometry of the ‘‘coupled'' meniscii would thus be analogous to the one of coupled ribbons with $\gamma_+/\gamma_-$ in place of $B_+/B_-$ as the governing parameter in Eq.~\ref{eq:main_sin}.

\begin{figure*}
\centering
\includegraphics[width = 1\textwidth]{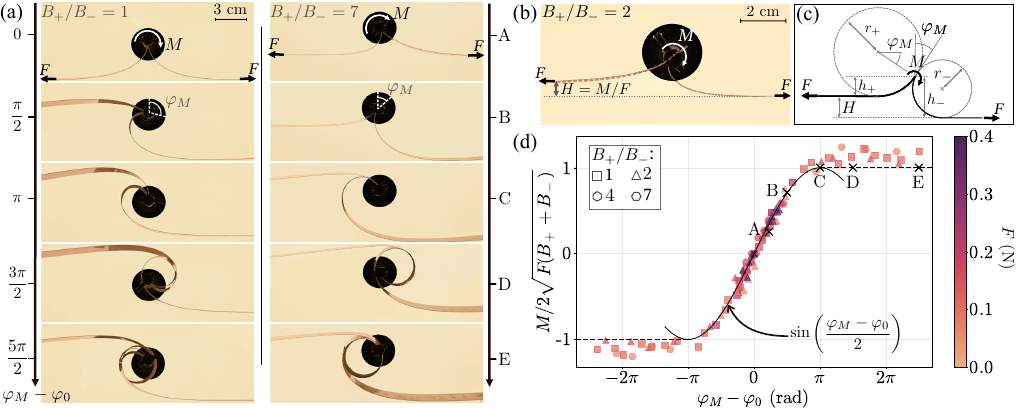}
\caption{Rotational response.  (a) Two coupled identical ribbons (left) and two ribbons with an asymmetry $B_+/B_- = 7$ (right) as the angle $\varphi_M - \varphi_0$ increases. The rescaled torque $M^* = M/2\sqrt{F(B_++B_-)}$ saturates beyond $\varphi_M - \varphi_0 = \pi$. (b) Two coupled ribbons with an asymmetry $B_+/B_-=2$ with an applied torque $M$. The dashed curves correspond to the predicted shape with two rescaled \textit{elastica} solutions. (c) Geometric approximation of the system with an added torque. (d) Master curve of the rotational behavior: rescaled torque $M^*$ versus the angular difference $\varphi_M - \varphi_0$ for various pulling forces (colorbar) and ratios of asymmetry (markers). The black line corresponds to Eq.~\ref{eq:torsion_model}, and the dashed lines to the predicted torque saturation. \corr{Note that on these figures $M$ and $\phi$ are counted positively in the clockwise direction}}
\label{fig:torsion}
\end{figure*}


\paragraph{Mechanical behavior} \corr{Coming back to elastic ribbons, a natural question is the rotational response of the junction. We now consider the addition of a point torque $M$ to the coupled end of the ribbons, fixing the junction point in translation while allowing its rotation. Two symmetric and coaxial forces $F$ are still applied to the free ends of the ribbons, and those ends are allowed to shift along the direction perpendicular to the force. We note the modified junction angle between the connected end of the ribbons and the direction perpendicular to the applied force as $\varphi_M$. When $M=0$, $\varphi_M$ reduces to the initial angle $\varphi_0$. Fig.~\ref{fig:torsion} (a) shows sequential pictures of the application of a torque on ribbons with $B_+/B_- = 1$ (left) and $B_+/B_- = 7$ (right), corresponding to experiments in Movies SI 3-4.} We introduce the dimensionless applied torque as $M^* = M/2\sqrt{F\left(B_+ + B_-\right)}$ as justified thereafter. The junction rotates in the direction of the applied torque with $\varphi_M - \varphi_0$ initially linear with respect to $M^*$ (A-B in Fig.~\ref{fig:torsion}). \corr{A shift perpendicular to the forces appears between the two free ends of the ribbons so as to balance the applied torque.} After reaching a threshold (C) that corresponds to $\varphi_M - \varphi_0 = \pm \pi$, the system cannot bear any larger torque. Increasing the angle further leads to the ribbons coiling onto themselves at a constant torque $M^* = \alpha$ with $\alpha$ slightly larger than one (D-E).

We now rationalize these observations. The governing equations~\ref{eq:def_system} and the geometric coupling still hold while the torque balance equation at the junction becomes $B_+\dot{\theta}_+ - B_-\dot{\theta}_- = -M$. The ribbons are therefore no longer aligned vertically although they still follow opposite directions. A distance $H=M/F$ measured perpendicularly to the pulling forces is observed between the two ribbons (Fig.~\ref{fig:torsion} (b)). The shape of the junction is still a combination of two partial classical solutions of the \textit{elastica} (Eq.~\ref{eq:full_sol_loop}) with appropriate scalings. A graphical construction method is used to obtain the shape by distancing the two asymptotic branches vertically by a distance $M/F$. Fig.~\ref{fig:torsion} (b) shows the solution overlaid on an experimental picture. The junction shape is thus prescribed by both the stiffness ratio $B_+/B_-$ and the length $M/F$.

We furthermore characterize the angular response $\varphi_M - \varphi_0$ to the applied torque $M$. Eqs.~\ref{eq:def_system} with the additional torque can be solved to obtain the following relation:
\begin{equation}
    \frac{M}{2\sqrt{F\left( B_+ + B_-\right)}} = \sin\left( \frac{\varphi_M - \varphi_0}{2} \right).
    \label{eq:torsion_model}
\end{equation}

\noindent 
The same result is obtained up to a factor $\sqrt{2}$ using a geometric approximation as shown in Fig.~\ref{fig:torsion} (c). Fig.~\ref{fig:torsion} (d) shows a great agreement of the model for experiments at various pulling forces made with samples of different stiffness asymmetry. \corr{Let us emphasize that this collapse is surprisingly universal even outside the linear regime. The stiffness asymmetry, expressed by the binding angle at rest $\varphi_0$, does not affect the global torsional behavior of the boundary layer (for example  the coiling always starts for an additional rotation of $\pm \pi$). We see that the two bending stiffnesses play a symmetric role in the rotational response.} Moreover, we can note that the rotational response is predominantly linear across a substantial range, essentially an interval of width $\pi$ centered on the rest angle $\varphi_0$. The system could thus be considered as a minimal visual torque sensor in this interval, especially considering the dependency of the stiffness with the square root of the pulling force.

Theory predicts the maximum admissible dimensionless torque to be unity, although we experimentally found a threshold value slightly larger than 1 which we attribute to friction in the system as in interleaved assemblies~\cite{Alarcon2016}. Coincidentally, the maximum torque $M = 2\sqrt{F(B_++B_-)}$ always occurs when the junction is rotated by half a turn $\varphi_M=\varphi \pm \pi$, at which point the curvatures of the ribbons become equal, regardless of $B_+/B_-$. Upon further rotation, the two ribbons develop extended contact and coil at a fixed torque with a radius determined by self-contact with a maximal value~$\sqrt{(B_+ + B_-)/F}$. This behavior is reminiscent of constant torque mechanisms \cite{dilly2023traveling}, e.g. power springs used in watchmaking based on the coiling of thin metal sheets \cite{saerens2022constant}.


\corr{Interestingly, these results can be easily extended to the case where the rods have a natural curvature on each side $\kappa_+$ and $\kappa_-$, with no assumption on their relative magnitude or sign. Considering no externally applied torque and an uniform curvature, \textit{elastic\ae} Eqs.~\ref{eq:def_system} are still valid, only the boundary conditions are modified. The torque balance at the junction becomes $B_+(\dot{\theta}_+ - \kappa_+) - B_-(\dot{\theta}_- - \kappa_-) = 0$ and can be rewritten into $B_+ \dot{\theta}_+ - B_-\dot{\theta}_- = B_+\kappa_+ - B_-\kappa_-$. In this form we recognize the previous problem for straight rods with an applied torque, and we can thus directly use Eq.~\ref{eq:torsion_model}. The junction angle $\varphi_\kappa$ depends on the asymmetry in both bending stiffness and natural curvature, given by:}

\corr{\begin{equation}
     \sin\left( \frac{\varphi_\kappa - \varphi_0}{2} \right) = \frac{B_-\kappa_- -B_+\kappa_+ }{2\sqrt{F\left( B_+ + B_-\right)}}
    \label{eq:curvatures}
\end{equation}}
\noindent
\corr{where $\varphi_0$ is obtained from Eq.~\ref{eq:main_sin}. The elastic junction of curved rods thus behaves similarly to the one of straight rods under the assumption of asymptotic pulling, with a correction on the junction angle based on their intrinsic curvature given by Eq.~\ref{eq:curvatures}.}


\paragraph{Measuring bending stiffness} We finally present a direct application of our results: a straightforward method to measure the relative bending stiffnesses of two materials through coupling. Let us consider two ribbons of unknown bending stiffnesses $B_a$ and $B_b$. One can easily determine the ratio of $B_a/B_b$ by binding the two ribbons so that one of their ends point towards the same direction, as done throughout this paper, and measuring the binding angle $\varphi_0$ while pulling on the other ends. Inverting Eq.~\ref{eq:main_sin}, we obtain $B_a/B_b = \left(1-\sin\varphi_0\right)/\left(1+\sin\varphi_0\right)$ independently of the force. The measurement is thus purely visual, with no required knowledge of or control over the applied force which makes it applicable at small scales. This approach is particularly valuable for heterogeneous materials where the classical formulas for the bending stiffness do not apply, such as composites and textiles \cite{goetzendorf2014bending, plaut2015formulas}. Moreover, pulling strongly on the ribbons allows to minimize the effects of geometrical defects and natural curvatures. We named this measurement as the $\lambda$-test in reference to the shape of the boundary layer.

To be accurately deployed, this method requires a force larger than $B/L^2$, where $L$ is the length of the ribbon. However the force should be lower than the adhesion force $\Gamma w$, where $\Gamma$ and $w$ are the adhesion energy and the width of the ribbon. In addition the junction should not yield plastically. The stress across the thickness of a bent ribbon scales as $\sigma \sim Et/\ell$ with $E$ the Young's modulus, $t$ the thickness and $\ell\sim \sqrt{B/F}$ the characteristic curvature. Considering a yield strength $\sigma_Y$, the maximal applicable force should thus be $ F_Y \sim \sigma_Y^2B/Et$ to ensure elasticity of the junction. \corr{Residual natural curvature perturbs the junction angle by $( B_-\kappa_- - B_+\kappa_+)/\sqrt{F(B_+ + B_-)}$ as seen from Eq.~\ref{eq:curvatures}, so that with an applied force larger than $\left(B_-\kappa_- - B_+\kappa_+\right)^2/\left(4\left(B_+ + B_-\right)\right)$ the effect of these curvatures is negligible}. Finally, to ensure the slenderness assumption, we should have $F < B/t^2$.

\begin{figure}
\centering
\includegraphics[width = \linewidth]{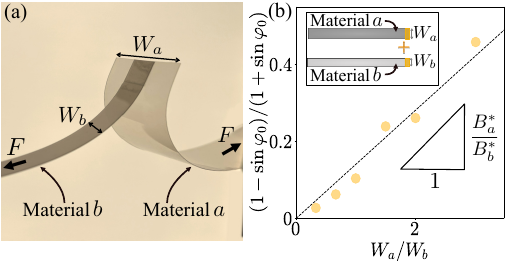}
\caption{The $\lambda$-test: measurement of relative bending stiffness. (a) Two coupled ribbons made of materials $a$ and $b$ (respectively 100 and 200 $\mu$m thick PET) and of widths $W_a$ and $W_b$. (b) $\left(1-\sin\varphi_0\right)/\left(1+\sin\varphi_0\right)$ as a function of $W_a/W_b$. The slope of the linear fit corresponds to the ratio of the bending stiffnesses per unit of width. The inset shows the geometry of the ribbons along their length.} 
\label{fig:FigApplicationMeasurement}
\end{figure}

A more important limitation remains in the sensitivity of the governing equation. Derivation gives its sensitivity: \corr{$d\varphi_0/d(B_+/B_-) = - \left(\sqrt{B_+/B_-}\left(B_+/B_- +1\right)\right)^{-1}$}, which tells us that the method is the most precise close to the symmetric case and worsens as the asymmetry increases. A first solution to this issue is to use the test in a cascade, starting with a known, relatively stiff material and trickling down over a range of materials, with a slight asymmetry only between each couples. Another solution, relevant for ribbons and plates only, is to use the fact that stiffness varies linearly with the widths of the ribbons $W_a$ and $W_b$. We thus denote their bending stiffness per unit width as $B_a^*$ and $B_b^*$. We fabricate pairs of ribbons with varying width ratios as illustrated in Fig.~\ref{fig:FigApplicationMeasurement} (a), and measure the corresponding angle at rest $\varphi_0$. Similarly, a linear relation is expected between $W_a/W_b$ and $\left(1-\sin\varphi_0\right)/\left(1+\sin\varphi_0\right)$ with a slope corresponding to $B_a^*/B_b^*$. Fig.~\ref{fig:FigApplicationMeasurement} (b) demonstrates the successful characterization of two reference polymer sheets using this method. Another convenient application of the ``$\lambda$-test" would be to test bound ribbons cut from the same sheet at different orientations, providing a direct method to characterize material anisotropy.


\paragraph{Concluding remarks} Our findings have revealed and characterized an asymmetric bending boundary layer commonly observed in various elastic systems. \corr{Notably, the rotational behavior of the junction displays a surprising universality}. While we experimentally considered ribbons, the results directly translate to other slender structures such as rods and plates. Unexplored limits of our work are the regimes beyond the assumptions of slenderness and elasticity, both caused by excessive forces. The lack of the former hypothesis introduces a stress distribution across the section, leading to shear and possible localizations like kinks. Exceeding elastic limits adds a complex interplay of the mechanical characteristics of the two bound objects based on the thresholds and types of plastic behaviors.

In the modeling of thin sheets it is common to neglect bending stiffness under large tensile loading, particularly for inflatable \cite{skouras2013computational, panetta2021computational} and tensile fabric structures \cite{bridgens2012form}, only missing details with vanishing small scales. The existence of the binding angle $\varphi_0$ as a scale-free geometric marker of the ratio of bending stiffnesses questions this practice. Moreover as in the example of Mylar balloons, we propose that such boundary layers at the connection of thin sheets could be used as an effective way to program motions in active materials. Finally, leveraging the independence of $\varphi_0$ from the tensile force, the $\lambda$-test provides a simple and reliable measurement of bending stiffness. We believe this tool could be of direct use to researchers and engineers working with soft or curved slender objects.

\bibliographystyle{ieeetr}
\bibliography{main}

\newpage
\onecolumn

This document contains information about the derivation of analytical results, methods for graphical construction and the presentation of a wetting system analogous to the coupled ribbons. We conclude with descriptions of the supplementary videos.

\section*{Derivation of analytical results}

We thereafter derive the junction angle $\varphi_0$ as a function of the stiffness asymmetry $B_+/B_-$ by solving two coupled rod equations (subsection A), and through a geometric approximation (B). We then solve the coupled equations with an additional applied torque for the rotational response $f$ defined as $\,dM/\,d\varphi_M$ as a function of $\left(\varphi_M, \varphi_0, B_+, B_-, F\right)$ (C), as well as using a similar geometrical approximation (D). We finally discuss the divergence and self-contact behavior of the rotational behavior (E).

\subsection*{A. Junction angle at rest using rod equations}

We consider two rods coupled at one end and pulled by two opposite forces of norm $F$. In the framework of the rods, with $s_{+,-} \in ]-\infty, 0]$ the arc lengths of the rods with $s_{+,-}=0$ at the junction, we write the two torque equilibrium equations with $\theta_+$ and $\theta_-$ \corr{the angle of the tangent to the rods with respect to the horizontal}:

\corr{
\begin{equation}
        \begin{cases}
        \begin{aligned}
            B_+\dfrac{\dd^2 \theta_+}{\dd s_+^2} - F\sin{\theta_+} = 0 \\
            B_-\dfrac{\dd^2 \theta_-}{\dd s_-^2} - F\sin{\theta_-} = 0
        \end{aligned}
        \end{cases}
    \label{eq:rod_system}
\end{equation}
}

\begin{figure}[ht]
\centering
\includegraphics[width = 0.55\textwidth]{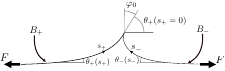}
\caption{Drawing of the two ribbons system in the framework of rod equations.  The difference in scaling of the two drawn ribbons corresponds to a stiffness asymmetry of $2.3$.} 
\label{fig:rod1}
\end{figure}

\noindent
\corr{Note that for the sake of symmetry the arc lengths are here oriented towards the junction point, and that the angles of the tangent of each rod $\theta_{+,-}$ are measured according to mirror-symmetric conventions and orientation.}
\corr{We also choose to count positively the junction angle $\varphi_0$ in the clockwise direction ($\varphi_0$ is positive on figure \ref{fig:rod1}).}

With the asymptotic condition  $\theta_{+,-}(s_{+,-} \to - \infty) \to 0$, we can multiply by $d\theta_{+,-}/ds_{+,-}$ and integrate the equations between $-\infty$ and any given $s_{+,-}$:

\begin{equation}
        \begin{cases}
        \begin{aligned}
            B_+\dfrac{1}{2}\left(\dfrac{\dd \theta_+}{\dd s_+}\right)^2 + F\left(\cos{\theta_+} - 1\right) = 0, \\
            B_-\dfrac{1}{2}\left(\dfrac{\dd \theta_-}{\dd s_-}\right)^2 + F\left(\cos{\theta_-} - 1\right) = 0.
        \end{aligned}
        \end{cases}
        \label{eq:b_integ_of_equilibrium}
\end{equation}
\noindent
At the junction point, both arc lengths correspond to $s_{+,-} = 0$. The balance of torques between the two rods writes:

\begin{equation}
    B_+\frac{\dd \theta_+}{\dd s_+}(s_+=0) - B_-\frac{\dd \theta_-}{\dd s_-}(s_-=0) = 0.
    \label{eq:b_torque balance}
\end{equation}
\noindent
Taking the square of Eq. \ref{eq:b_torque balance} and introducing it into Eqs. \ref{eq:b_integ_of_equilibrium}, we get:

\begin{equation}
    FB_+ \left(\cos\theta_+\left(s_+=0 \right) - 1 \right) - FB_- \left(\cos\theta_-\left(s_-=0 \right) - 1 \right) = 0.
\end{equation}
\noindent
With the coupling equation being \corr{$\theta_-\left(s_-=0 \right) = \pi - \theta_+\left(s_+=0 \right)$} and having defined the junction angle as $\varphi_0 = \pi/2 - \theta_+\left(s_+=0 \right)$, we get:
\begin{equation}
\corr{2FB_+ \left(\sin\varphi_0 - 1 \right) + 2FB_- \left(\sin\varphi_0 + 1 \right) = 0}.
\end{equation}
The expression for the junction angle is thus:

\begin{equation}
    \sin\varphi_0 = \frac{B_+ - B_-}{B_+ + B_-}.
    \label{eq:junction_angle}
\end{equation}.
\subsection*{B. Junction angle at repose using a geometric approximation}

\begin{figure}[h!]
\centering
\includegraphics[width = 0.4\textwidth]{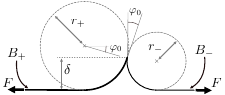}
\caption{Drawing of the two ribbons system with idealized curvatures at the junction. The difference in curvatures at the junction corresponds to a stiffness asymmetry of $1.5$.} 
\label{fig:scaling1}
\end{figure}

One can actually obtain the result of Eq.~\ref{eq:junction_angle} in an approximate but simpler system. We idealize the shape of the rods near the junctions as arcs of circle of radii $r_+$ and $r_-$ as illustrated on Fig.~\ref{fig:scaling1}. The torque balance at the junction writes:

\begin{equation}
    \frac{B_+}{r_+} = \frac{B_-}{r_-}.
    \label{b:torque_balance}
\end{equation}

\noindent Furthermore, the point of junction of the two rods corresponds to the intersection of the circles. The junction angle $\varphi_0$ is therefore also the angle between the direction of the pulling force with the line that links the center of the two circles, see Fig.~\ref{fig:scaling1}. By construction, we express the height of the junction $\delta$ as a function of the circle radii and the junction angle:

\begin{equation}
\begin{aligned}
 \delta & = r_+ - r_+\sin \varphi_0, \\
 & = r_- + r_-\sin \varphi_0.
     \label{b:height_simple}
\end{aligned}
\end{equation}

\noindent Inputting Eqs.~\ref{b:height_simple} in Eq.~\ref{b:torque_balance} gives the junction angle:

\begin{equation}
    \sin\varphi_0 = \frac{B_+ - B_-}{B_+ + B_-}.
    \label{b:result_theta0}
\end{equation}
\noindent
We can go one step further and get the scaling of the height. To do so, we write the torque balance of each individual rod at the junction point. The pulling force indeed acts with a lever arm $\delta$ on the junction point, the torque balance for the stiffer rod gives:

\begin{equation}
    \frac{B_+}{r_+} = F\delta,
\end{equation}

\noindent and for the less stiff one:

\begin{equation}
    \frac{B_-}{r_-} = F\delta.
\end{equation}

\noindent Inputing either equation in Eqs.~\ref{b:height_simple} with~\ref{b:result_theta0} gives us the scaling of the height:


\begin{equation}
    \delta = \sqrt{2F\frac{B_+B_-}{B_+ + B_-}} .
\end{equation}

\subsection*{C. Rotational stiffness using rod equations}

\begin{figure}[h]
\centering
\includegraphics[width = 0.55\textwidth]{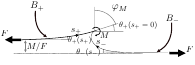}
\caption{Drawing of the two ribbons system with an external torque applied at the junction in the framework of rod equations. The difference in scaling of the two drawn ribbons corresponds to a stiffness asymmetry of $2.3$.} 
\label{fig:rod2}
\end{figure}

We aim to determine the rotational stiffness $\,dM/\,d\varphi_M$ as a function of $\left(\varphi_M, \varphi_0, B_+, B_-, F\right )$ of the system as schematized in Fig.~\ref{fig:rod2}, \corr{where the applied point torque $M$ is counted positive in the clockwise direction.} We consider two coupled rod equations as in Eqs.~\ref{eq:rod_system}. Accounting for the applied torque $M$, the coupling condition now writes:
\corr{
\begin{equation}
    B_+\frac{\dd \theta_+}{\dd s_+}(s_+=0) - B_-\frac{\dd \theta_-}{\dd s_-}(s_-=0) = -M.
\end{equation}}

\noindent Using the first integration of the rod equations expressed in Eqs.~\ref{eq:b_integ_of_equilibrium}, we get:

\begin{equation}
    \sqrt{2B_+F}\sqrt{1 - \cos\theta_+(s_+=0)} -
    \sqrt{2B_-F}\sqrt{1 - \cos\theta_-(s_-=0)} = -M.
\end{equation}

\noindent Geometrically, we have $\theta_-(s_-=0) = \pi - \theta_+(s_+=0)$ and $\varphi_M = \pi/2 - \theta_+(s_+=0)$. We can thus write the previous equation as a function of $\varphi_M$ only:

\begin{equation}
    \sqrt{1 - \sin\varphi_M} - \sqrt{\frac{B_-}{B_+}} \sqrt{1+ \sin\varphi_M} = -\frac{M}{\sqrt{2FB_+}}.
    \label{eq:stiff_intermediaire}
\end{equation}

\noindent Note that $M/\sqrt{2FB_+}$ is the usually expected dimensionless torque in the \textit{elastica} framework. Next, we can use the relationship Eq.~\ref{eq:junction_angle} demonstrated in subsection (A) to express \corr{$B_-/B_+ = \left(1-\sin\varphi_0\right)/\left(1+\sin\varphi_0\right)$}. We can thus write:

\begin{equation}
    \sqrt{1-\sin \varphi_M}\sqrt{1+\sin\varphi_0} - \sqrt{1-\sin\varphi_0}\sqrt{1+\sin\varphi_M} = -\frac{M}{\sqrt{2FB_+}}\sqrt{1+\sin\varphi_0}
\end{equation}

\noindent
We rewrite the term on the right noting that $1 + \sin\varphi_0 = 2/\left(1+B_-/B_+\right)$ using Eq.~\ref{eq:junction_angle}. We then obtain:

\begin{equation}
    \frac{1}{\sqrt{2FB_+}}\sqrt{1+\sin\varphi_0} = \frac{1}{\sqrt{FB_+}\sqrt{1 + B_-/B_+}}.
\end{equation}

\noindent The latter equation therefore expresses a dimensionless applied torque generalized to any ratio of asymmetry $B_-/B_+$.
Considering the term on the left, it is possible to demonstrate trigonometrically that for any $(x,y) \in [0, \pi]^2$:
\corr{
\begin{equation}
    \sqrt{1 -\sin x}\sqrt{1 + \sin y} - \sqrt{1 + \sin x}\sqrt{1 - \sin y} = 2\sin\left(\frac{y - x}{2}\right)
\end{equation}
}


\noindent Using this relationship in Eq.~\ref{eq:stiff_intermediaire}, we finally get the torsional behavior of the system:
\begin{equation}
    \sin\left(\frac{\varphi_M - \varphi_0}{2}\right) = \frac{M}{2\sqrt{F \left( B_+ + B_- \right) }}.
    \label{eq:solution1_torsional equation}
\end{equation}

\noindent From this equation, we get the entire behavior observed experimentally as shown in Fig. 4 of the main paper.

\subsection*{D. Rotational stiffness using a geometric approximation}

\begin{figure}[h!]
\centering
\includegraphics[width = 0.4\textwidth]{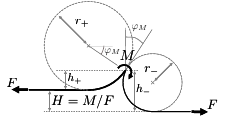}
\caption{Drawing of the two ribbons system with an external torque applied with idealized curvatures at the junction. The difference in curvatures at the junction corresponds to a stiffness asymmetry of $1.5$.} 
\label{fig:scaling2}
\end{figure}

We get a  solution similar to Eq.~\ref{eq:solution1_torsional equation} using the geometric approximation used in subsection (B). We draw the approximation in Fig.~\ref{fig:scaling2}.

\noindent
First, the geometric approximation through torque balances gives:

\begin{equation}
    H + h_+ = h_-,
\end{equation}

\begin{equation}
    -M/F + r_+\left(1 - \sin \varphi_M\right) = r_-\left(1 + \sin \varphi_M\right).
    \label{eq:scaling2_geo1}
\end{equation}
\noindent
Then, from the torque balance of each individual rod we note that $h_- = B_-/(r_-F)$ and $h_+ = B_+/(r_+F)$. Geometry  also gives \corr{$h_- = r_- \sqrt{1+\sin \varphi_M}$ and $h_+ = r_+ \sqrt{1-\sin \varphi_M}$}. Inserting those relations into Eq.~\ref{eq:scaling2_geo1} leads to:

\begin{equation}
    \sqrt{1 - \sin\varphi_M} - \sqrt{\frac{B_-}{B_+}} \sqrt{1+ \sin\varphi_M} = -\frac{M}{\sqrt{FB_+}}.
    \label{eq:scaling_bad}
\end{equation}

\noindent
This equation is the same, up to a factor of $\sqrt{2}$, as Eq.~\ref{eq:stiff_intermediaire}. We then follow the steps of subsection (C) to obtain a similar relation than Eq.~\ref{eq:solution1_torsional equation} having used the geometric approximation.

\subsection*{E. Torque limit and coiling behavior}

From Eq.~\ref{eq:solution1_torsional equation}, we easily see that the absolute value of the rescaled torque cannot exceed a value of unity and that this maximum is reached for $\varphi_M - \varphi_0 = \pm \pi$. At this point, extended contact between the two rods appears as we have an equality of the curvatures. From the Eqs.~\ref{eq:b_integ_of_equilibrium}, we get the expression of the curvatures as \corr{$1/\dot{\theta}_+ = \sqrt{2B_+/(F(1 - \sin \varphi_M))}$ and $1/\dot{\theta}_- = \sqrt{2B_-/(F(1 + \sin \varphi_M))}$}.
\noindent
The equality of the curvatures is therefore equivalent to:

\begin{equation}
     \frac{1 + \sin\varphi_M}{1 - \sin \varphi_M} = B_-/B_+ 
\end{equation}
\noindent
We recognize here the similarity of this equation with Eq.~\ref{eq:junction_angle}. The only values of $\varphi_M$ that fulfill this equation are $\varphi_0 \pm \pi$. Therefore, the point at which the limit in torque of the system is reached corresponds as well to the one at which the rods have a contact with the same curvature. This is the mathematical condition for developing an extended contact, as observed when coiled further. The fact that it corresponds to the maximal torque sustained by the system is a striking coincidence.

\section*{Shape of the junction using graphical constructions}

We describe here how to obtain the full shape of the coupled ribbons starting from the known solution of the \textit{elastica} presented in the main paper. We first discuss the case of coupled ribbons under tension only, and then present the consequence of an additional applied torque at the junction.

\subsection*{F. Shape of the ribbons under tension}

We consider two ribbons of stiffnesses $B_+$ and $B_-$ pulled in opposite directions by forces of norm $F$. Out of the two possible length scales, we arbitrarily choose $\sqrt{B_+/F}$ to fix the size of our graphical constructions. We remind the reader of the shape of the solution of the \textit{elastica} which presents one loop and is pulled asymptotically with $\theta(s \to -\infty) = 0$ and $\dot{\theta}(s \to -\infty) = 0$, scaled correspondingly to the stiffer ribbon:

\begin{figure}[h!]
\centering
\includegraphics[width = 0.9\textwidth]{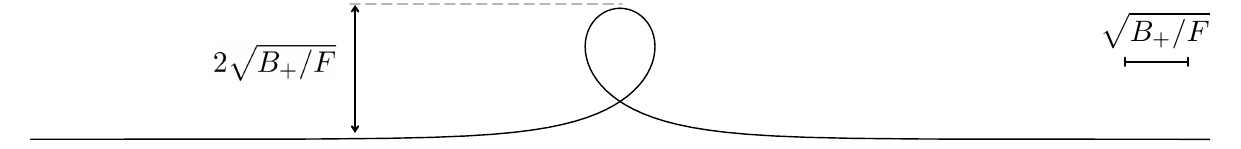}
\end{figure}

\noindent
The solution is arbitrarily truncated on the left and on the right. Let us then consider an example with two ribbons that exhibit an asymmetry such as $B_+/B_- = 3$. We can plot two solutions with the corresponding scalings. For the sake of clarity, we omit half of the loops.

\begin{figure}[h!]
\centering
\includegraphics[width = 0.9\textwidth]{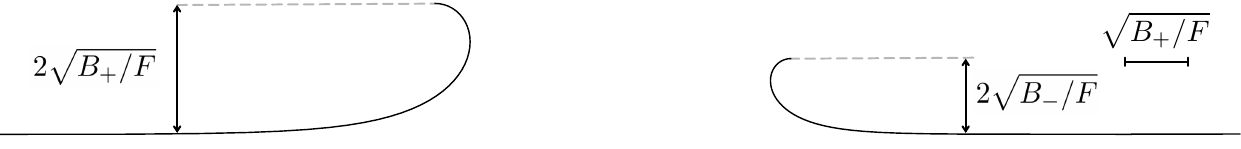}
\end{figure}
\noindent
The two asymptotic branches follow opposite directions by construction of the problem. They are as well necessarily aligned to ensure the torque balance of the coupled ribbons system. Let us now close the gap between the two solutions.

\begin{figure}[h!]
\centering
\includegraphics[width = 0.9\textwidth]{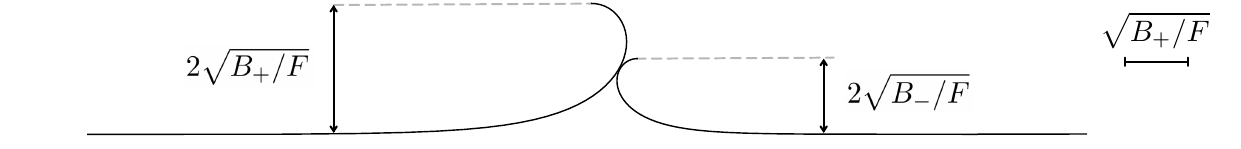}
\end{figure}
\noindent
Matching tangents and angles are found between the curves at the first point of contact, meaning that the two junction conditions are met ($\theta_- = \pi - \theta_+$, and $B_+\dot{\theta}_+ = B_-\dot{\theta}_-$, see section III for detailed notations). We can therefore highlight in red the shape of coupled ribbons for a stiffness asymmetry of $B_+/B_- = 3$:

\begin{figure}[h!]
\centering
\includegraphics[width = 0.9\textwidth]{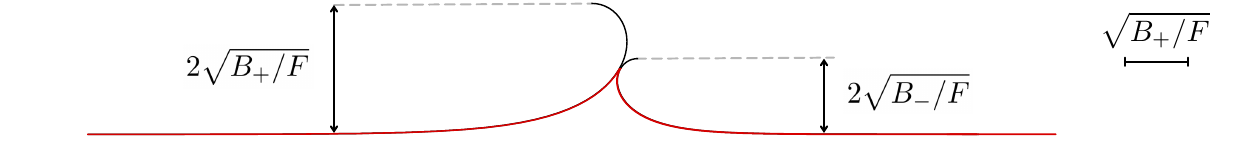}
\end{figure}
\noindent
Geometrically, the solution is therefore governed by the square root of the ratio of the bending stiffnesses $\sqrt{B_+/B_-}$. We present in Fig.~\ref{fig:5_quelques_exemples} the graphical constructions for several other stiffness asymmetries.

\subsection*{G. Shape of the ribbons with an external torque}

We consider here the shape of the coupled ribbons with an applied torque $M$ at the junction. To ensure global torque balance, this requires the two asymptotic directions to be separated by a distance of $F/M$. Otherwise, the corresponding graphical construction still follows the same concepts. We start again with two partial solutions to the \textit{elastica} with a stiffness asymmetry of $B_+/B_- = 3$, and consider here as an example a torque $M = \sqrt{FB_+}/2$:

\begin{figure}[h!]
\centering
\includegraphics[width = 0.9\textwidth]{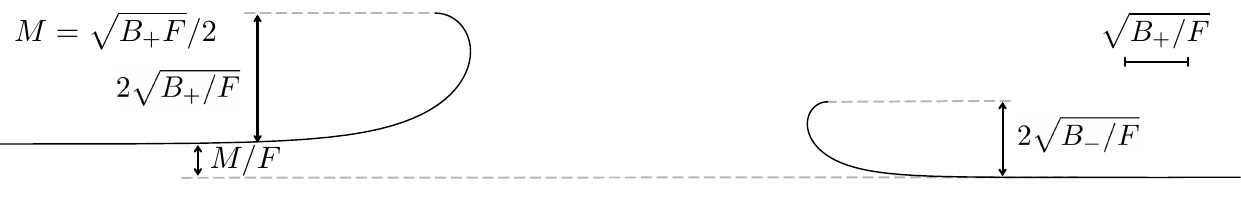}
\end{figure}

\noindent
Closing the gap between the two half loops, the first point of intersection still corresponds to one that fulfills the junction conditions ($\theta_- = \pi - \theta_+$, and through the introduction of the distance $M/F$ we also have $B_+\dot{\theta_+} = B_-\dot{\theta_-} + M$).

\begin{figure}[h!]
\centering
\includegraphics[width = 0.9\textwidth]{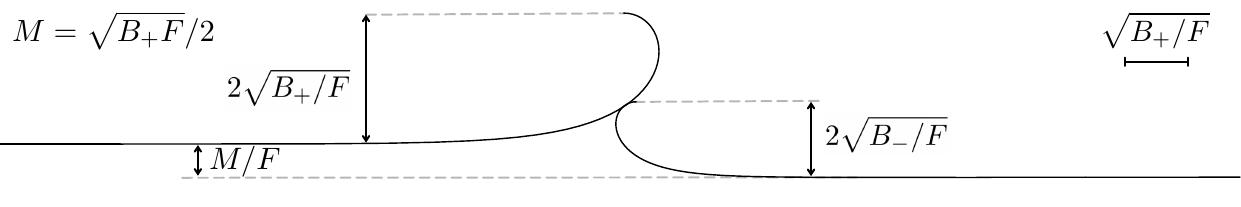}
\end{figure}
\noindent
We thereafter highlight in red the actual shape of the coupled ribbons.

\begin{figure}[h!]
\centering
\includegraphics[width = 0.9\textwidth]{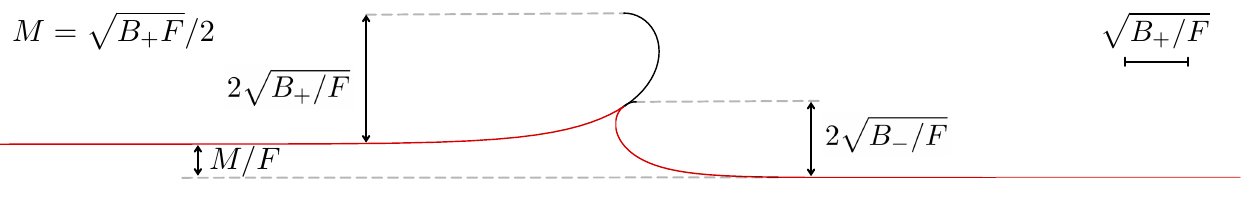}
\end{figure}
\noindent
Geometrically, the solution is governed both by the square root of the ratio of the stiffness asymmetry $\sqrt{B_+/B_-}$ and the length scale associated to the applied torque and tensile force $M/F$. We present in Fig.~\ref{fig:p2_quelques_exemples} the corresponding shapes for a given asymmetry $B_+/B_- = 3$ while increasing the ratio of $M/F$. Note that at given torques which correspond to $M = 2\sqrt{B_-/F}$ and $M =-2\sqrt{B_+F}$, we need to flip one of the solution, with no consequence on the construction. Furthermore, we reach a point at which we have an equality of the curvatures on each side at the limit $M = \pm2\sqrt{F(B_+ + B_-)}$. Further increasing the torque leads to a situation at which the intersection point do not fulfill both the geometrical compatibility and the torque balance conditions. Coincidentally, it corresponds to the point at which extended contact between the two rods is observed.

\newpage
\section*{Analogy with coupled liquid meniscii}

The form of the \textit{elastica} equation is shared with the equation of a two-dimensional liquid meniscus. The family of curves based on the coupling of two coupled ribbons as described in this paper is therefore as well the solution to a system of coupled meniscii. We present here this analogous system.\\
\noindent
Let us first write the equation that governs a 2D liquid meniscus:
\begin{equation}
    \gamma\dfrac{\dd^2 \theta}{\dd s^2} - \rho g\sin{\theta} = 0,
\end{equation}

\noindent with $\gamma$ the surface tension of the fluid, $\rho$ its density, and $g$ the gravitational constant. The characteristic radius of curvature of a meniscus is the  capillary length $\sqrt{\gamma/(\rho g)}$, in place of $\sqrt{B/F}$ for ribbons. We can then consider two meniscii with different surface tension $\gamma_+$ and $\gamma_-$, that are nonetheless associated to fluids of equal density.  One experimental way of doing so would be to consider surfactants to differentiate in surface tension two air-liquid interfaces of one fluid and to separate them with a plate, a configuration reminiscent of a Langmuir trough, here with rotational freedom.

\begin{figure}[h!]
\centering
\includegraphics[width = 0.7\textwidth]{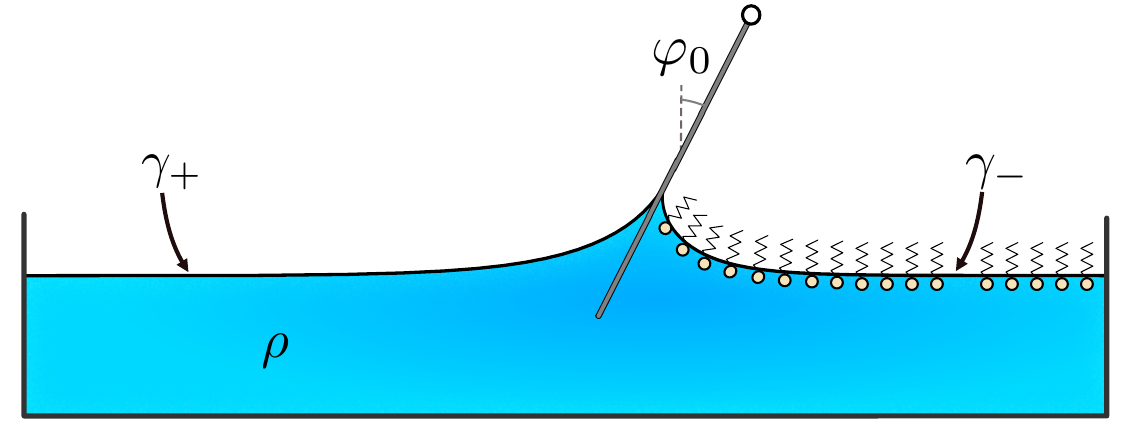}
\caption{System of two meniscii coupled through a freely rotating, stiff plate. On the right, surfactant compounds reduce the natural surface tension of the fluid. The shape of the coupled meniscii is the same as the one of coupled ribbons under tension.} 
\label{fig:analogy_menisque}
\end{figure}

Fig.~\ref{fig:analogy_menisque} shows a drawing of our imaginary 2D system. A fluid sits in a container that is much larger than the addition of the two capillary lengths $\sqrt{\gamma_+/(\rho g)}$ and $\sqrt{\gamma_-/(\rho g)}$. Near the middle of the container, a stiff plate separates the surface of the air-liquid interface in two parts. The weightless plate is hanging from the air, fixed in translation and free to rotate. We assume that the fluids completely wet the plate. We consider surfactants on one side of the plate only, therefore reducing the surface tension on this side to $\gamma_-$. Through the torque balance of the plate, we expect the meniscii to match in height.

The shape of the meniscii should therefore be described by the same family of curves described in this paper about coupled ribbons, with the ratio of $\sqrt{\gamma_+/\gamma_-}$ as the governing parameter rather than $\sqrt{B_+/B_-}$. While we believe this analogy interesting in itself, we shall note that the system presented here appears hardly practical.

\clearpage
\paragraph{Movie S1.} Two identical ribbons made of 200 \si{\micro \meter} thick PET ($B_+/B_- = 1$) are pulled apart by hand. One can observe the self-similar behavior of its shape through the transformation. 
\paragraph{Movie S2.} Two ribbons made of 200 \si{\micro \meter} thick PET, the one on the left being $0.5$ \si{\centi \meter} wide and the one on the right $1.5$ \si{\centi \meter} wide ($B_+/B_- = 3$), are pulled apart by hand. One can observe the self-similar behavior of its non-symmetric shape through the transformation.
\paragraph{Movie S3.} Two identical ribbons made of PET ($B_+/B_- = 1$) are pulled at a given force $F = 10$ \si{\milli\newton}, and a torque is applied by hand at their junction. After the junction angle reaches $\pi$, the system starts coiling itself due to self-contact. The experiment stops when the ribbons buckle out-of-plane, a phenomenon not studied here.
\paragraph{Movie S4.} Two ribbons made of PET, the one on the left 100 \si{\micro \meter} thick and the one on the right 200 \si{\micro \meter} ($B_+/B_- = 7$), are pulled at a given force $F=5$ \si{\milli\newton}, and a torque is applied by hand at their junction. After the junction angle reaches $\varphi_0 + \pi$, the system starts coiling itself due to self-contact until it buckles out of plane.

\end{document}